\documentclass[18pt,preprint]{aastex}
\begin{document}
\title{Spectroscopic observations of a sample of dwarf spirals galaxies. I.Oxygen abundances}
\author{A.M. Hidalgo-G\'amez}
\affil{Departmento de F\'isica, Escuela Superior de F\'{\i}sica y Matem\'aticas, IPN, U.P. Adolfo L\'opez Mateos, 
C.P. 07738, Mexico city, Mexico}
\author{D. Ram\'{\i}rez-Fuentes}
\affil{Instituto de Astronom\'\i a, UNAM, Ciudad Universitaria, Aptdo. 70 264,
C.P. 04510, Mexico City, Mexico}
\author{and J. J. G\'onzalez}
\affil{Instituto de Astronom\'\i a, UNAM, Ciudad Universitaria, Aptdo. 70 264,
C.P. 04510, Mexico City, Mexico}

\begin{abstract}

The oxygen abundances for four dwarf spiral galaxies have been determined using long-slit spectroscopy.  The abundances of these galaxies have not  previously reported in the literature. Several H\,{\sc ii} regions were detected in each galaxy. The electronic temperature method could be used only in one region because of the lack of the auroral lines of oxygen or sulfur. Therefore, four different semi empirical methods were used in the abundance determinations and a weighted-average abundance is obtained. The abundances of three of the galaxies are sub solar for all the regions, one of them having a very low metallicity. Only three H\,{\sc ii} regions in the fourth galaxy show slightly over solar abundances. Three of the H\,{\sc ii} regions might have a PN embedded. A comparison with the oxygen abundances among different types of late-type galaxies (BCG, LSBG, dI, Sm) is made. The conclusion is that all of them show the same range in metallicity. Also, the log(N/O) is similar, showing a secondary behaviour for abundances larger than $8.2$ dex in spite of the morphological type of the mother galaxy.


\end{abstract}

\section{Introduction}

The concept of dwarf galaxy was firstly introduced in a publication of the Kapteyn Astronomical Laboratory just few months after the astronomical community accepted the idea of galaxies outside the Milky Way. By 1936 there were at least three dwarf (or under luminous) galaxies in Local Group (Hubble 1936). In the next two decades another nine dwarf galaxies were discovered in the Local Group, but also in other groups of galaxies, as M81 and M101 (Hodge 1958). It was claimed that dwarf galaxies was discovered in the Virgo Cluster.  Reaves (1956) classified four galaxies of the Virgo cluster as dwarf spiral galaxies, although he said that ``the existence of these objects has not been established'' and their existence ``is inferred by analogy with the dwarf variants of normal elliptical and spirals''. But, when in the late 50`s de Vaucouleurs reorganised the Hubble sequence, the terminology dwarf spiral galaxy almost disappeared from the literature. Because of their scarcity, they were included in the new types of spiral galaxies, as Sd and Sm. Few years later, Nilson (1973) used this terminology, as dS, to classify $12$ of the galaxies in the Uppsala General Catalog. From this time to 1995 most of the references to dwarf spiral galaxies were to deny their existence (e.g. Edmunds \& Roy 1993). Actually, Drinkwater \& Hardy (1989) worried about the missing dwarf spiral galaxies of the Virgo Cluster. Sandage et al. (1985), from their study of the Luminosity Function in the Virgo cluster, claimed that although there were previous evidence of small Sm galaxies, none Sa to Sc dwarf galaxy was ever discovered. Finally, Schombert et al. (1995) discovered six small galaxies with distinct bulge and disc components, low surface brightness, blue colours and flocculent spiral patterns which they classified as early-type dwarf spiral galaxies. 

The properties of the late-type dS (or small size Sm galaxies) were discussed deeply in Hidalgo-G\'amez (2004; hereafter HG04). A total of $111$ galaxies classified as late-type spirals, with optical radius smaller than $5$ kpc and absolute magnitude lower than $-18$ were found. Their colours were bluer than normal-size Sm but their Star Formation Rates were very low (Reyes-P\'erez \& Hidalgo-G\'amez 2008). Very few of the dS galaxies in the Hidalgo-G\'amez's sample might be located inside clusters or group of galaxies but the number of barred galaxies is higher that for normal Sm galaxies, probably due to their small sizes which make them very unstable dynamically. Few of these dS galaxies show a clear spiral pattern. Instead, torn off arms are observed, not well defined and most of the time only one arm is detected. From the abundances of nine dS galaxies it was concluded that the oxygen abundances were similar or smaller than the values for the LMC (see Table 4 in HG04). 

Hidalgo-G\'amez \& Olofsson (2002) determined that the integrated oxygen abundances of Im compared to Sm galaxies from the classical work of McCall et al. (1985) is smaller by $0.2$ dex. One can think that such differences are not very large. Actually, they are within $1$$\sigma$ uncertainties of the measurements, as stated in Table 4 in Hidalgo-G\'amez \& Oloffson (2002). In recent years, the abundance has been determined for another nine dS galaxies and the results were that most of them have low, or very low, abundances, lower than 12+log(O/H) = $8.2$. Therefore, the range of abundances encompasses by the so far studied from a sample of $18$ dS galaxies is very similar to these of Im and normal Sm galaxies (See Table 9 and table 4 in Hidalgo-G\'amez 2004). This is not expected because of the, a priori,  different Star Formation Histories in these types of galaxies. It is supposed that the Star Formation (SF) in spiral galaxies is continuous due to the dynamical spiral arms, while it proceeds on burst, separated by $10^8$ years, in Im galaxies (Carroll \& Ostlie 2007). Therefore, the metal content should be larger in the former than in the latter if the same IMF works at both type of galaxies. Moreover, these two types seems to have different environments (HG04), which might influence the Star Formation Rates. Then, it might be of the outmost importance to understand such coincidence in the metallicity range.  If this hold when more galaxies are included, different conclusions can be attained: the spiral arms of the Sm and dS might not be dynamical and they do not influence the SF; or the IMF differs for Im galaxies, favouring the existence of a large number massive stars, which produce and release heavier elements as oxygen. Or all these galaxies are really the same kind of galaxies, and the classification scheme is so much detailed. Or some other differences in the evolution of the late-type galaxies.  Therefore, a larger sample of abundance for late-type galaxies must be considered before any further speculation on their evolution.

The main goal of this investigation is to increase the number of dS galaxies with spectroscopic information and metallicity determination. Therefore, four newer dS galaxies were observed and their oxygen abundances determined, for which not previous values have been reported. In a companion paper (Hidalgo-G\'amez et al., AJ submitted; hereafter paper II), the existence of oxygen abundance gradients is studied. In the following, the basic information for the galaxies in the sub sample studied here is presented. Also, a summary of this information is presented in Table 1. 

UGC 5242 is a barred galaxy with an optical size of $4.86$ kpc and located at a distance of $24.7$ Mpc (HG04). Despite to be at the Vorontson-Velyaminov catalogue (1974), it is not classified as a peculiar or interacting galaxy. It does not show disturbances in its morphology, although the spiral structure is not very well defined as clearly seen in the optical images (those obtained by one of the authors or those provided by SLOAN). As several H\,{\sc ii} regions are clearly seen in the H$\alpha$ image this galaxy is a perfect candidate to study the oxygen abundance.

UGC 5296 is one of the smallest galaxies in the sample of HG04 with an optical radius of only $3.02$  kpc. The absolute magnitude is also small, $-15.08$ if a distance of $20.28$ Mpc is adopted. Despite its important amount of gas, $2 \times$ 10$^8$ M$\odot$, only five H\,{\sc ii} regions are clearly distinguished in an H$\alpha$ image of the galaxy. No spiral pattern can be distinguished in the galaxy in the H$\alpha$, V or R images despite being classified as Sm (Nilson 1973). 

UGC 6205 is, on the contrary, one of the largest galaxies in the dS sample with an optical radius of $4.99$ kpc, but with a low gas mass, only $3\times10^8$ M$\odot$, and a low luminosity, M$_B$ = $-16.5$ at a distance of $18.9$ Mpc. The spiral structure of this galaxy does not show any disturbance in all the bands (V, R and H$\alpha$) we observed. Actually, this galaxy display the best spiral structure out of the four galaxies studied here, with several H\,{\sc ii} regions visible in the H$\alpha$ image taken previously to the spectra. 

UGC 6377 is one of $12$ galaxies classified as dwarf spirals in the Uppsala General Catalog (Nilson 1973). Very little information is available for this galaxy. If a distance of $27.3$ Mpc is considered, M$_B$ is about $-15.5$. Again, the H$\alpha$ image showed several H\,{\sc ii} regions but not clear spiral structure. 

The structure of the paper is the following: the description on the acquisition of the data and the reduction processes is presented in the next Section. In Section 3 a discussion on the abundance methods is presented. The values of the oxygen abundances for each H\,{\sc ii} regions are presented in Section 4 while a discussion on them is presented in Section 5. Conclusions are given in Section 6.

\section{Observations and data reduction}

First of all, it has to be said that the sample of galaxies observed and presented here did not suffer any selection before observations, not due to the number of H\,{\sc ii} regions detected or a more clear (or fuzzy) spiral pattern. The only selection criteria were the coordinates and the hour angle at the precise moment of the observations. Actually, this was on purpose in order not to include any bias which might make these galaxies different from the rest of the galaxies in Table 1 of HG04. In fact, they cover a large range in properties, as seen in Table 1, being a very representative sample of dS galaxies.

The spectra on which this investigation is based on were acquired on 2002 March 10~-~11 with the 2.1 m telescope of the Observatorio Astron\'omico Nacional at San Pedro M\'artir (Mexico). The Boller \& Chives spectrograph was used with a $300$ line mm$^{-1}$ grating blazed at 5000 \AA. The detector was a SITe3 CCD with $3$ $\micron$ pixels in a 1024$\times$1024 format. The slit length was $5$ arcmin and the width was $170$ $\micron$m, subtending $\approx$ 2 arcsec on the sky, and yielding a spectral resolution of $7$\AA. The pixel scale along the spatial direction was $1.05^{''}$. The full spectral range observed was $3500-6900$ \AA. During the two nights the weather conditions were good but not completely photometric. The seeing was about $1.5^{''}$ most of the time. A summary of the observing conditions and log of observations is shown in Table 2. No correction for differential refraction was performed, as the airmasses were, in all cases, smaller than $1.5$ but for the last slit position of UGC 6377. Actually, the integration time of the two last positions in this galaxy was small due to the rapid increase of the airmass. The orientation of the slits were east-west and parallel to each other for the galaxies UGC 5242 and UGC 6377 and north-south for the galaxies UGC 6205 and UGC 5296. 

The reduction of the data was performed with the VISTA software package, an automatic procedure developed at Lick Observatory at mid-eighties and often updated by one of the authors (J.J. G\'onzalez). Bias and sky twilight flat-fields were used to calibrate the CCD response. Several exposures of a given slit position facilitated the removal of cosmic rays. He-Ar lamps were used for the wavelength calibration and geometric deflections corrections. The spectra were corrected for atmospheric extinction using the San Pedro M\'artir tables (Schuster \& Parrao 2001). The flux calibration was performed using several standard stars, observed each night at different airmasses. The flux accuracy was of the order of 5$\%$. Finally, the number and extension of the H\,{\sc ii} regions were determined. To do so, the forbidden oxygen line [O\,{\sc iii}]$\lambda$5007~\AA~ was used. A final spectra for each H\,{\sc ii} region was created summing up all the rows where this line was observed.

The next step is to correct these spectra for absorption and extinction. The usual procedure for the absorption correction is to increase the equivalent width of $H\beta$ by $2$~\AA~ (e.g. McCall et al. 1985). However, the differences in the intensity of the absorption corrected line ratios and the non-absorption corrected ones were of $2\%$ when no clear absorption features are presented in the spectra (Hidalgo-G\'amez 2007). Such differences are of the same order as the error due to the Poisson noise and, therefore, not significant uncertainties were introduced if the absorption correction is not performed. The extinction correction was performed using the extinction coefficient, defined as 
\begin{equation}
 C_{\beta} = {1 \over f(\lambda)} ln {I(H\alpha)/I(H\beta) \over 2.86} 
\end{equation}

where $2.86$ is the theoretical Balmer decrement at $10,000$ K for case B of recombination (Brocklehurts 1971), $f(\lambda)$ is the Whitford modified extinction law (Savage \& Mathis 1979), and $I(H\alpha)/I(H\beta)$ is the observed  H$\alpha$/H$\beta$ ratio. When this ratio is smaller than the canonical value $2.86$, the extinction coefficient was set to zero, as usual. 

The intensities of the lines detected in each spectrum were measured with two different softwares: VISTA and the subroutine ALICE in MIDAS. The first one detected and measured the lines by an automatic procedure which know the wavelength of few spectral lines, while a more ``manual-visual'' procedure is used in ALICE. The software at VISTA could be used only for two of the galaxies studied here, UGC 5242 and UGC 5296. All the lines were measured twice with ALICE. If the two values differed by more than $50\%$, a third measurement was carried out in order to obtained the most reliable value of the intensity. This was normally the case for low signal-to-noise (S/N) lines. A normalisation to the intensity of H$\beta$ was performed, as usual.
The uncertainty for each line was obtained from the quadrature of the uncertainties in the reduction procedure ($\sigma_{r}$), in the extinction correction ($\sigma_{e}$) and in Poisson noise ($\sigma_{c}$). From this set of intensities the abundances of oxygen can be determined.

\section{On the oxygen abundance determination}

A total of $32$ H\,{\sc ii} regions were detected in the four galaxies observed: $12$ regions in UGC 6205, $9$ in UGC 6377, $7$ in UGC 5242 and only $4$ in UGC 5296. The regions were named with a letter indicating the slit position and a number increasing from  south to north (or east to west in UGC 6205 and UGC 5296). In two of the regions of UGC 5242 only the H$\alpha$ line was detected and in one of the UGC 6205 regions the H$\beta$ line was absent. Therefore, the abundances can be determined in only  $29$ of the H\,{\sc ii} regions observed. The auroral line of oxygen at $4363$\AA~ was not detected in any except one of the regions and, therefore, the abundances cannot be determined with the standard method (Aller 1984; Osterbrock 1989) for $28$ of the H\,{\sc ii} regions.  

\begin{figure}
\centering
\includegraphics[width=8cm]{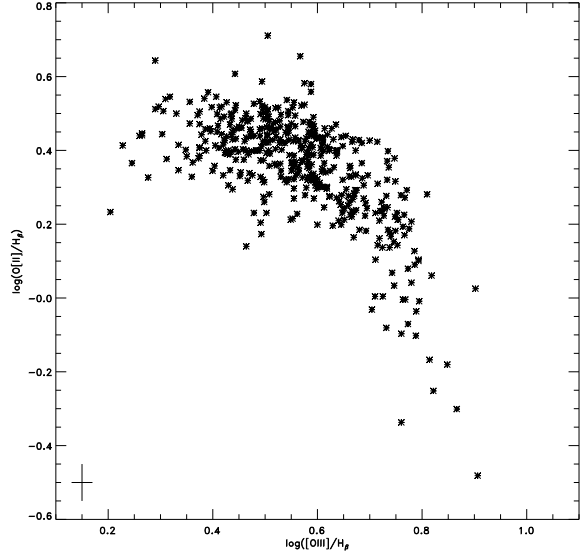}
\caption{The [OII]/H$\beta$ vs. [OIII]/H$\beta$ intensities from a sample of 438 star-formation galaxies. The line is the best fitting to the data points. The largest dispersion is between $0.6 <$ log([OIII]/H$\beta$) $< 0.8$, which is about $35\%$. At the bottom-left part of the figure the cross represents the typical errorbar of the data-points} 
\label{fig1}
 \end{figure}

In any case, the most important caveat concerning this set of data is the lack of the forbidden oxygen doublet at $\lambda3726-\lambda~3729$\AA~ (hereafter, $\lambda$3727) due to the low efficiency of the spectrograph at the blue end. Without the intensity of these lines, the oxygen abundances cannot be determined for more than half of the regions, $16$ out of the total $29$ H\,{\sc ii} regions.  Therefore, it might be interesting to find out any way to obtain information on the intensity of the [O\,{\sc ii}] doublet. This could be through the relationship between the intensities of [O\,{\sc iii}]$\lambda$5007\AA~ and [O\,{\sc ii}]$\lambda$3727\AA. Such relationship can be obtained directly from photoionization models (Mart\'{\i}n-Manj\'on et al. 2008) and it has been used as a diagnostic diagram (e.g. Baldwin et al. 1981). In the present investigation, we used the relationship obtained from a sample of $438$ star forming galaxies by Kniazev et al. (2004). Such relationship, which is shown in figure ~\ref{fig1}, follows the equation  

\begin{equation}
log{[OII] \over H\beta} = -2.82~ log({[OIII] \over H\beta})^2 + 2.39~ log({[OIII] \over H\beta}) - 0.0542
\end{equation}

with a rather large dispersion, of around $35 \%$. Kobulnicky et al. (1999) proposed a similar procedure for high redshift galaxies, for which the oxygen line [O\,{\sc iii}]$\lambda$5007 is not detected. They did not give any equation to compare with  but they obtained a similar plot in both, shape and dispersion, to the one in figure ~\ref{fig1}. Although the dispersion of such relationship is large, a value for the intensity of the [O\,{\sc ii}]$\lambda$3727 can be obtained and, therefore, the abundances for all the H\,{\sc ii} regions can be determined with confidence.   
 
The next step is to derive the oxygen abundances. As previously said, the so-called standard method cannot be used due to the lack of the auroral lines in the spectra. Therefore, the semi empirical methods should be used instead. Four different semi empirical methods were used here to determine the oxygen abundances. In Hidalgo-G\'amez \& Ram\'{\i}rez-Fuentes (2009), the accuracy of two of these methods were studied. They were focussed on the R$_{23}$ (Pagel et al. 1979; McGaugh 1994) and the $P$ method (Pilyugin 2000; 2001; Pilyugin \& Thuan 2005) because their sample did not have the intensities of the nitrogen lines. Although the new calibration of the $P$ method gives values of the metallicity closer to those obtained with the standard method in general, the differences with the R$_{23}$ method are not that impressive and, therefore, it cannot be concluded that $P$ method's values are preferred. For a deeper discussion on this issue the reader is referred to the original paper by Hidalgo-G\'amez \& Ram\'{i}rez-Fuentes (2009). The main problem with these two semi empirical methods is that there are bi-valuated. This means that there are two zones of metallicity for the same value of $[OII]+[OIII]$ intensity.  For the $R_{23}$ method they are the so-called low-metallicity branch and high-metallicity branch (Edmunds \& Pagel 1984). In the case of the $P$ method, there are two different equations for the abundance determination: one for abundance values lower than $8.2$ and other for values larger than this. Therefore, we must establish an ``a priori'' criterion in order to determine in which of these 
two regions the abundance of the H\,{\sc ii} region under studied is. The most widely criterion used is based on the log([N\,{\sc ii}]/[O\,{\sc ii}]) ratio. When this ratio is higher than $-1.0$, the region is considered of high metallicity. Given the uncertainties in the [O\,{\sc ii}] intensities described before, this ratio could be lower (or higher) than $-1.0$ and the region is on the high (or low) metallicity branch. Moreover, for those regions where the abundances can be obtained from other methods, they will be used as a guess on the metallicity branch. In this sample, the nitrogen line at $\lambda 6583$ \AA~ is  detected only in half of the regions. Therefore, the location on the metallicity branch is not straightforward for half of the sample. 

The other two methods used in the determination of the oxygen abundances in these H\,{\sc ii} region are based on the nitrogen line. The relation between the oxygen abundance and the [N\,{\sc ii}]/H$\alpha$ ratio (hereafter the $N2$ method) was recently calibrated by Denicol\'o et al. (2002). Also, the calibration proposed by Pettini \& Pagel (2004) based on the $log ({[NII]/H\alpha \over [OIII]/H\beta})$ ratio and the oxygen abundances (hereafter, $N3$) is considered. The precision of these two methods was not studied by Hidalgo-G\'amez \& Ram\'{\i}rez-Fuentes (2009) and therefore, we are going to relay on the total uncertainties determined by Denicol\'o et al. and Pettini \& Pagel, respectively. The typical uncertainty is, in both cases, about $0.2$ dex.

A final concern about the semi empirical methods is their dependence on other parameters besides the oxygen abundances such as the ionization temperature (Olofsson 1997), or the nitrogen content (Stasi\'nska 2008).  The dependence  on the ionization parameter has been studied, among other, by McGaugh (1991). Some information on the ionization parameter can be obtained through the [O\,{\sc ii}] vs. [O\,{\sc iii}] space and photoionization models. Following the models by Garnett (1999), the ionization parameters for all the regions studied here except two (U05c1 and U96a2) agree with a value of $log~U = -3$, as well as an ionization temperature of $40,000$~-~$50,000$ K. Moreover, we used the calibrations given by Kobulnicky et al. (1999) for the determination of the abundances with the R$_{23}$ method.  Such calibrations were provided by McGaugh based on his photoionization models, which take into account the ionization parameter through the $y$ parameter. It is also important to mention that the $N3$ might be saturated near the solar abundance (Erb et al. 2006) and, therefore, for larger values the $N3$ method is not reliable. This is not a real problem here because only two regions show values of $8.5$ dex with this method, and they are similar to the other three measurements (U05a1) or very much different (U96a1).

All together, four different abundances are available for half of the sample of H\,{\sc ii} regions and, therefore, a reliable value can be inferred. For the other half of the sample only two values of the abundances can be obtained and therefore, the final abundance will be less accurate. The extinction-corrected intensities of the lines for each region are presented in Tables 3 to 6 along with their uncertainties, the extinction coefficient and the signal-to-noise (S/N) ratio in the [O\,{\sc iii}]$\lambda$5007 line. Three of the galaxies in the sample, UGC 5242, UGC 5296 and UGC 6377, have spectral information in the SDSS data-base \footnote {Funding for the SDSS and SDSS-II was provided by the Alfred P. Sloan Foundation, the Participating Institutions, the National Science Foundation, the U.S. Department of Energy, the National Aeronautics and Space Administration, the Japanese Monbukagakusho, the Max Planck Society, and the Higher Education Funding Council for England. The SDSS was managed by the Astrophysical Research Consortium for the Participating Institutions}. However, these SDSS data are not of better quality than those presented here, with the [O\,{\sc ii}]$\lambda$3727, [N\,{\sc ii}]$\lambda$6583 as well as the [O\,{\sc iii}]$\lambda$4363 lines are absent for all of the SDSS spectra. Using the same procedure described here an oxygen abundance can be obtained and they were compared with those obtained at the SPM-OAN telescope.

\section{Oxygen abundances in dwarf spiral galaxies}
 
The oxygen abundances were determined from the extinction-corrected intensities of the lines for each region. 
A different oxygen abundance is obtained with the different methods for each region. With all of these abundances, a weighted-average abundance value is determined. In order to do so, the results from Hidalgo-G\'amez \& Ram\'{\i}rez-Fuentes (2009) were considered, giving more weight to the metallicity determined with the $P$ method for low and high abundances zones if the ionization ratio ([O\,{\sc iii}]/[O\,{\sc ii}]) is larger than $1$ but a similar weight for those values when this ratio is smaller than $1$ as well as in the so-called ``turn-around'' region.  A dispersion from this weighted-average value is determined and tabulated in Table 7 along with the averaged metallicity value.
 
\subsection{UGC 6205}

As previously said, a total of $12$ H\,{\sc ii} regions were observed in this galaxy in three different slit positions. They were orientated N-S, with a rotator angle of $270°$. Slit position named $a$ was located at the canonical coordinates of the galaxy, therefore these regions are the most central ones. Slit position named $b$ was moved towards the north about $24$ arcsecond and the slit named $c$ was moved towards the south about $13$ arcsecond from the central position, in both case without changing the rotator angle. The H$\beta$ line was not detected for one of the regions, therefore we can get abundance values for $11$ regions. The intensities and their uncertainties of 
all the lines detected are shown in Table 3. The [N\,{\sc ii}]$\lambda$6583 line was detected in only four of these eleven regions: U05a1, U05b2, U05b4 and U05c3. Therefore, the abundances of these four regions can be determined with a total of four different methods. For them, the log ([N\,{\sc ii}]/[O\,{\sc ii}]) ratio as well as the $N2$ and $N3$ abundances  will be used to determine the metallicity branch of the regions. These values are presented in Table 7. The log([N\,{\sc ii}]/[O\,{\sc ii}]) ratio has a very high value for U05a1, and the $N2$ and $N3$ abundances are also very high. Therefore, they are consistent each other. On the contrary, this ratio is slightly larger than $-1$ for region U05c3, but the $N2$ and $N3$ abundances are larger than $8.2$ dex and, therefore the high-metallicity branch values from the R$_{23}$ and $P$ methods were preferred. For U05b2 the ratio is lower than $-1$, although with large uncertainties, while their $N2$ and $N3$ abundances lie between the low-metallicity and high-metallicity branch. Actually, the low and high abundance values from the R$_{23}$ and the $P$ methods for this region are very similar. The most troublesome region is U05b4,  because the abundances determined with the $N2$ and the $N3$ are larger than $8.2$ dex while the log (N/O) ratio is clearly lower than $-1.0$. An abnormal intensity of the nitrogen line could not be the explanation for such difference because the nitrogen intensity is very similar to the values obtained in other H\,{\sc ii} regions in this galaxy, as U05b2. In
any case, the high-metallicity values are very similar to those from the $N2$ and $N3$ in U05b4 and, therefore, they are preferred.

The metallicities of the other seven regions where the nitrogen line at $\lambda$6583\AA~ is not detected are more difficult to determine. One of the reasons for the lack of detection of this line might be the low signal-to-noise (S/N) of their spectra. Actually, all the regions without the [N\,{\sc ii}]$\lambda$6583 line but one (U05c1) have lower S/N than those regions where the nitrogen line is detected. Therefore, it cannot be concluded that all these regions have low-metallicity values. In any case, both values of the abundances (low- and high-metallicity) were determined for all the regions from both methods. From an inspection of these values, it is clear that the oxygen abundance of U05c2 is closer to the value inferred from the high-metallicity branch because the low-metallicity values are lower than $7.3$ dex. For the rest of the regions, their galactocentric distances will be used to determine the metallicity branch. For it, it will be assumed that those regions located towards the center will have higher abundances than those located towards the edge. As regions U05a2 and U05a3 are located inside the central $700$ pc, it is expected that they will have high-metallicity abundances and those values will be considered here. On the contrary, U05b1, U05b3 and U05b5 are outside $1.5$ kpc and the low metallicity branch values will be preferred. Finally, for U05c1 and U05c2, which are located in the intermediate part of the galaxy, the high metallicity abundance values will be preferred. 

This situation described above is the most typical for spiral galaxies (Skillman et al. 1996). As only very few galaxies shows no differences in the oxygen abundance with the galactocentric distance (e.g. NGC 1313, Walsh \& Roy 1997) we thought this will be the most likely scenario for UGC 6205. In any case, there is a possibility that this galaxy shows a very shallow gradient, with all the abundances from the high-metallicity branch. Then, the values of U05b1, U05b3 and U05b5 will be of $8.55$, $8.42$ and $8.63$ dex, respectively. The main caveat is that these abundances are higher than the values for the intermediate distance regions, which will make the gradient to be negative for the innermost part of the galaxy and positive for the outermost regions. To our knowledge, this has not been observed in any spiral galaxy before. In any case, a deeper discussion on this effect will be held in paper II and here the low-metallicity values for these three regions are considered and stated in Table 7.  

The regions with the largest abundances are U05c1, U05a3, U05a2, and U05a1. Meanwhile U05a2 and U05a3 are located very close to the center of the galaxy, U05c1, and U05a1 are at intermediate distances, between $1$ and $1.6$ kpc. Therefore, their abundances might be anomalously high. The most interesting feature in the spectrum of U05a1 is the large value of the [N\,{\sc ii}]/H$\alpha$ ratio, of $0.8$. This might be indicative of shocks but also some other phenomena, such as planetary nebulae (PN). As the sulfur lines were not visible and the [OII]$\lambda$3727 line could not be measured, the discrimination between the H\,{\sc ii} and the PN nature cannot be done. Concerning U05c1, the high [S\,{\sc ii}]/H$\alpha$ ratio of the region, of $0.55$ it is very interesting and therefore, the existence of shocks inside this region should be considered. Since the nitrogen line is absent in U05c1, it is not very likely the existence of a PN inside it. The shock waves could be due to the spiral structure, which can be clearly seen in the H$\alpha$ images of this galaxy. Actually, the [S\,{\sc ii}]/H$\alpha$ ratio for almost all the regions in this galaxy is very high, pointing out to shocks induced by the spiral structure. 

From the results showed in Table 2 UGC 6205 is a typical spiral galaxy, with the central regions having solar abundances and those at the outskirts showing low metallicity when some of the regions of high metallicity are not considering as pure H\,{\sc ii} regions.

\subsection{UGC 6377}

A total of $3$ slit positions were observed for this dwarf spiral galaxy orientated E-W, with a rotator angle of $0°$.  As no H$\alpha$ image of was available for this galaxy, the coordinates were used to position the first slit, named $b$. Slit named $a$ was positioned towards the east about $10$ arcsecond and the slit named $c$ was moved towards the west about $13$ arcsecond from the center, without changing the rotator angle. In these three slit positions, a total of $9$ H\,{\sc ii} regions were detected. In  one of them, U77a2, the forbidden auroral oxygen line at $4363$\AA~ is clearly seen. Therefore, the abundances can be determined using the so-called standard method (e.g. Allen 1984; Osterbrock 1989). The value of the electronic temperature of the high ionization zone, from the [O\,{\sc iii}] intensity, as well as the oxygen and nitrogen abundances are shown in Table 8. Also, the log(N/O) and the electron temperature of the low-ionization region were determined and showed in this table. About the T$_e(O^+)$ determination, the reader is referred to Hidalgo-G\'amez \& Ramirez-Fuentes (2009) for a detailed discussion on it. Despite the high electron temperature, $20,000$ K, similar to the values with the less metallicity galaxies known (Papaderos et al. 2008), the oxygen abundance is not very low, only of $7.8\pm0.1$ dex. This behaviour could be due to an abnormally high intensity of the [O\,{\sc ii}] line or could be real because there is not a linear correlation between the electron temperature and the oxygen abundances (Hidalgo-G\'amez, in preparation).  When, and according to figure ~\ref{fig1}, the lowest value of the intensity of the [O\,{\sc ii}] line is considered, an abundance of $7.4$ dex is obtained, which is lower than expected because this region is located inside $1$ kpc from the center.  The nitrogen abundance is very low as well as the log(N/O) ratio. In spite of these values, the abundances determined with any of the semi empirical methods are not so low but intermediate (see Table 7), in agreement with the first value obtained of $7.8$ dex. Finally, it is quite interesting that this region follows very well the relationship between nitrogen and oxygen abundances found for a sample of irregular galaxies (see figure 3 in Hidalgo-G\'amez \& Olofsson 2002). 

For the other eight regions, the abundances has to be determined with the semi empirical methods due to the lack of the auroral lines. The nitrogen line [N\,{\sc ii}]$\lambda$6583 is detected in almost half  of the regions, including U77a2. The values of the log(N/O) ratio, as stated in Table 7, indicate that two of them, U77a1 and U77b1, might have high-metallicity values while the other two, U77a2 and U77c2, have low-metallicity ones. This is confirmed by their $N2$ and $N3$ abundances for each region (see Table 7). These results are quite odd in the sense that the H\,{\sc ii} regions with the highest abundances are located at the outskirts of the galaxy and this is not the expected situation for spiral galaxies, where the most external regions have  lower metallicities than the internal, as previously said. As the standard method provides a low abundance for U77c2 and this region is quite close to the center of the galaxy, there are several possibilities to explain the high abundances of U77a1 and U77b1. The most likely possibility is that there is an extra mechanism which increases the intensities of some of the lines and, therefore, the abundance appears artificially higher than it really is.  From an inspection of their spectra, it is remarkable than the oxygen line [O\,{\sc i}]$\lambda$6300\AA~ is clearly detected in the spectrum of U77a1, with a [O\,{\sc i}]/H$\alpha$ ratio of $0.1$. Although this value is high, it can be due only to photoionization (Stasi\'nska 1990). On the other hand, and according to the diagnostic diagrams (see Section 5 and figures ~\ref{fig2}, ~\ref{fig3} and ~\ref{fig4}) this region is at the edge of the photoionized regions locus and therefore, shocks might be playing a role in the metallicity determination. Moreover, the intensity of the nitrogen line at $\lambda$6583\AA~ is very high, of $0.43\pm0.2$, compared to the average value for normal spiral galaxies, of about $0.11\pm0.1$ (McCall et al. 1985). The oxygen line [O\,{\sc i}]$\lambda$6300\AA~ is not detected in U77b1. In this case, the [S\,{\sc ii}]+[S\,{\sc ii}]/H$\alpha$ is very high, larger than $1$ . Moreover, the $[NII]/H\alpha$ ratio is of $0.3\pm0.2$. Both values might indicate  the existence of an extra ionization mechanism (e.g. shocks, Dopita \& Sutherland 1995). One explanation for the large values of the nitrogen intensities and the suspicion of shocks is that these are really planetary nebulae (PN) instead of H\,{\sc ii} regions. Another possibility is that these regions have low abundances despite of their $N2$ and $N3$ abundances. The low-metallicity abundances for U77a1 and U77b1 are $7.6$ and $7.8$ dex, respectively. These values agree very well with the abundance gradient determined in Paper II (see figure 2 in Paper II).  Therefore, the low-metallicity abundances might be the most likely values. In the following, the abundances of these two regions will be taken with care until new observations were performed and confirm the pure H\,{\sc ii} region nature.  

The [N\,{\sc ii}]$\lambda$6583 line is not detected in the other five regions. As U77a2 is located at $1$ kpc from the center with a low oxygen abundance, the agreement was that all the regions beyond this distance will be of low metallicity as well. Only one region is located inside $1$ kpc, U77b2, and it will be considered as of high metallicity, with a value of $8.4 \pm 0.1$ dex. In any case, the abundances are similar or slightly lower than the values for the LMC (e.g. Russell \& Dopita 1990). 

As previously said, an spectrum of UGC 6377 can be retrieved from the SDSS data-base. The quality of this spectrum is lower than those presented here, with the lack of the [O\,{\sc iii}]$\lambda$4363 and [N\,{\sc ii}]$\lambda$6583 lines as well as [O\,{\sc ii}]$\lambda$3727. The abundances, determined with the R$_{23}$ and $P$ methods are of $7.8 \pm 0.15$ dex or $8.45 \pm 0.1$ dex, for the low- and high-metallicity branch, respectively. As this galaxy is not that far away (only 27.3 Mpc) that the subtended angle were smaller than the SDSS fiber-size (of about 3''), this is not an integrated spectra, but more likely might correspond to the central part of the galaxy, because it is the region with the largest surface brightness. Therefore, the location might be similar to the one corresponding to U77b2 and the high metallicity branch value is preferred.

\subsection{UGC 5296}

The number of the H\,{\sc ii} regions in this galaxy is quite few. A total of only five regions were detected in the H$\alpha$ image (Reyes-P\'erez \& Hidalgo-G\'amez 2008). One of them is located to the  far east of the galaxy and no spectrum was obtained for it, while the other four regions were aligned north-south and those are the spectra presented in this investigation. Their spectra were obtained with only one slit position, due to their alignment, being the northeaster one, at $38$ arcsec from the center, named $a$ and the Southend one, $d$, at $22$ arcsec.  

The intensities of the lines were measured with both softwares, ALICE-MIDAS and VISTA. The values obtained are presented at Table 5. For H$\beta$ only one row is presented because the normalised intensity of this line is always $1$. The VISTA software detected the [N\,{\sc ii}]$\lambda$6583\AA~ line in all the regions, while only in two are clearly presented according to the visual inspection: U96a2 and U96a3, which are the regions with the highest S/N. As VISTA works as an automatic procedure for the line detections, it might be possible that the nitrogen line in the other two regions are not real, but only upper limits. Actually, the intensity of this line in U96a1 is the highest in the sample, but no line was detected in a visual inspection.  

Another issue to be noticed from the line intensities presented in Table 5 is that the line intensities measured by VISTA and by ALICE are very similar except for region 
U96a1. A reason for such difference is that this is the region with the lowest S/N in the sample and therefore, they were very difficult to measure with any of the softwares. 

The values of the log([N\,{\sc ii}]/[O\,{\sc ii}]) ratio are listed in Table 7.  Both regions U96a2 and U96a3 are clearly of high-metallicity because the values of this ratio with both, VISTA and ALICE, set of intensities are higher than $-1$. Moreover, the $N2$ and 
$N3$ abundances are larger than $8.2$ for both regions and both set of intensities. Therefore, they both will be considered as of high-metallicity. It is very interesting to see that the oxygen line at $\lambda~6300$\AA~ is detected in both regions. Moreover, the [O\,{\sc i}]/H$\alpha$ ratio and the [S\,{\sc ii}]/H$\alpha$ are very large, especially for U96a2, with values of $0.3\pm0.1$ dex and $0.45\pm0.1$ dex, respectively. On the contrary, the [N\,{\sc ii}]/H$\alpha$ is normal to the values obtained in other Sm galaxies, for both regions. This could be an indication of the existence of shocks inside this region, and an explanation for its large $R_{23}$ and $P$ abundances. 

The VISTA value for the log([N\,{\sc ii}]/[O\,{\sc ii}]) ratio in U96a4 is lower than $-1$, indicating probably a low metallicity value. The $N2$ and $N3$ VISTA abundances are of $8.26$, just on the limit between the high and low metallicity regions. The S/N ratio of this region is similar to U96a2, where [N\,{\sc ii}]$\lambda$6583\AA~ is clearly visible. Moreover, H$\gamma$ is detected in U96a4. Therefore, the lack of the nitrogen line is not due to the low S/N but to a real absence due to low metallicity. Moreover, this region is located at about $2$ kpc from the center and then, the low-metallicity values of the R$_{23}$ and $P$ abundances  are preferred. 

The most troublesome region is U96a1. The VISTA value of the log([N\,{\sc ii}]/[O\,{\sc ii}]) ratio is very high, but the $N2$ and $N3$ abundances are of $8.5$, probably because of the saturation of the $N3$ method at solar abundances (Erb et al. 2006). As previously said, the VISTA values of the line intensities are very different from those measured with ALICE. Moreover, the former intensities are very odd; its intensity of the [O\,{\sc iii}]$\lambda$5007 line is the lowest for all the regions in the sample but the [N\,{\sc ii}]$\lambda$6583 is very high. Correction for dispersion might be important in the determination of such odd metallicity. The main caveat is that all the H\,{\sc ii} regions of this galaxies were detected in only one single slit-position (or angle), then if dispersion correction (or the lack of it) is the problem, it might affect to all the values of the other three H\,{\sc ii} regions as well. In particular, to the values of U96a4 because it is also very far from the center of the galaxy (and the center of the slit). As the other regions do not present any peculiar intensities, we think that the lack of dispersion correction can be ruled out as the origin of the odd intensities (and oxygen abundances) of U96a1. Then, the galactocentric distance will be useful for deciding the metallicity branch. Actually, this is the most  outside region, at almost $3$ kpc from the center. With such galactocentric distance, values of the abundances as high as $8.9$ dex are not very likely. Therefore, the low-metallicity values are preferred. Although the metallicity determine with the low-metallicity branch, of $7.4$ dex for the average metallicity, is lower than the abundance for the H\,{\sc ii} region SDH323 in M101 (Kennicutt et al. 2003), this value is similar to those of low metallicity irregular galaxies as GR8, DDO 168 and DDO 167 (Hidalgo-G\'amez \& Olofsson 2002).

Due to its quite remarkable low metallicity of this H\,{\sc ii} region, we would like to explore other possibilities. The first one is take into account the values obtained from the VISTA software. The averaged oxygen abundance is $8.0$ dex, which is a normal value compared to the metallicities of other H\,{\sc ii} regions in Sm galaxies (McCall et al. 1985).  Another possibility is being U96a1 a distant galaxy in itself.  Along with the low metallicity is the fact that it is very far from the center, at $2900$ pc. Moreover, it is easily detected in H$\alpha$ but not in V or R. In spite of this, the receding velocity obtained from the spectrum is very similar among all the H\,{\sc ii} regions, including U96a1. Therefore, it will be considered as part of the galaxy.

There is also an spectrum of this galaxy in the SLOAN data-base, although the metallicity cannot be determined because no intensity for H$\beta$ line is reported. Moreover, the uncertainties in the H$\alpha$ intensity are about $90\%$.   

As can be seen, the metallicity of UGC 5296 follows the typical behaviour of normal spirals with a high-metallicity core, in this case the values are about solar, and low abundances at the edges. It is interesting to notice that the agreement in the metallicity values obtained from the VISTA set of intensities and the ALICE-MIDAS one is very good except for U96a1. 

\subsection{UGC 5242}

A total of $7$ H\,{\sc ii} regions were observed in this barred galaxy in five slit-positions orientated E-W (rotator angle $0°$), parallel each other. In two of them, located at the far east of the galaxy (at more than $25$ arcsec from the center), only the H$\alpha$ line was detected. The slit position named $d$ is located about the center of the galaxy, while $19$ arcsec towards the east is slit position named $c$ and $8$ arcsec towards the west is slit position named $e$. From these three positions, there are oxygen abundances determined for five regions. Again, the intensities of the lines were determined with two softwares, ALICE-MIDAS and VISTA. It is interesting to notice important differences in the line intensities between them. The main one is the lack of detections for region U42d1 with VISTA, while a normal, complete set of line were detected with ALICE-MIDAS. No explanation could be found for such situation, therefore only a set of line intensities was considered for this region. In addition to this, and as in UGC 5296, VISTA detected the [N\,{\sc ii}]$\lambda$6583 line in all the other four regions, while only in three out of five regions this line was detected using ALICE-MIDAS. Two of these regions, U42c2 and U42e2, have low S/N ratio and the VISTA [N\,{\sc ii}]$\lambda$6583 intensities are very low. Therefore, it will be considered as noise detection and the average abundance from the VISTA values in Table 7 were obtained without them. Finally, there are differences in almost of the line intensities between the two set for U42c1, which is the region with the largest S/N ratio. In spite of this, the abundaces from the two set of intensities are very similar for this region except for those obtained with the $N2$ method.

As for the previous galaxies, the log(N[\,{\sc ii}]/[O\,{\sc ii}]) ratio and the $N2$ and $N3$ abundances were used to determine the metallicity branch. The values of the log(N[\,{\sc ii}]/[O\,{\sc ii}]) ratio are stated in Table 7 for both set of intensities. It is remarkable than the VISTA values are much lower than the ALICE-MIDAS ones except for U42e1.  Actually, three of the values of the log(N/O) are very low, lower than $-1.9$. The log([NII]/[OII]) values indicate low metallicity for all the H\,{\sc ii} regions.  Using the low-metallicity values, they range between $7.7$ and $8.1$ dex, except two $N2$ VISTA values. Moreover, the agreement among the abundance values obtained with VISTA software and ALICE-MIDAS ones are very good for all the regions in UGC 5242. 

There are two spectra of this galaxy in the SDSS data-base but only one of them is retrievable. As this galaxy is at $20$ Mpc, the intensity of the [O\,{\sc ii}]$\lambda$3727 line is not available, and therefore, equation 2 is used. The value of the log(N/O) is $-1.11 \pm 0.2$ dex and the abundances determined with the $N2$ and $N3$ are lower than $8.3$ dex. Then, the low-metallicity branch values are preferred and an averaged abundances of $8.0$ dex are obteined with them. This value is very close to the one for U52d1. As for UGC 6377, the spectra will correspond likely to the central part of the galaxy. 

This galaxy does not follow the canonical situation for spiral galaxies. There is no real enhancement of the metallicity at the central part of the galaxy (as said, slit named $d$ is very close to the center of the galaxy). Instead, the abundances are very similar wherever their galactocentric distance. It resembles irregular galaxies, where the differences in the oxygen abundances among the H\,{\sc ii} regions do not follow any pattern, if they exist (Hidalgo-G\'amez et a. 2001b). Another situation might be a very shallow gradient, as typical of barred galaxies (Martin \& Roy 1994).

\section{Discussion}

One crucial question about the data presented here is how accurate these values are. This is important because they have been determined with semi empirical methods, which as discussed in Section 3, have several and important caveats and dependences. Nevertheless, the used of semi empirical method in the abundance determination in spiral galaxies is very common (e.g. Skillman et al. 1996). Moreover, the [O\,{\sc ii}] intensities are determined from a statistical approach. Therefore, it is important to check the consistency of the results.

Concerning the abundance values presented here, the first thing that can be notice from Table 7, is that the Z$_{23}$ abundances are always higher than the abundances from the other methods for more than half of the sample. The difference between the metallicity determined with the R$_{23}$ and any other method could be up to $0.2$ dex. This is probably related with the excitation parameter $P$, as stated by Pilyugin (2003). If those values with differences larger than $1.5~\sigma$ are not considered in the determination of the average metallicity, the values are still not very much different from those reported in column 7 of Table 7. Similar results are obtained if only the regions with the highest S/N are considered. Moreover, the dispersion values of the abundances for all the regions except eight (including the VISTA abundances) are $0.1$ or smaller. Such small value of the dispersion indicates a large consistency between the different abundance estimations. Some of the regions with the larger dispersion are located in UGC 5296 and UGC 5242, for which the VISTA abundances have larger dispersion than the ALICE-MIDAS ones. In some case, the discrepancy is between pairs of values: those estimates with the R$_{23}$ and $P$ methods are consistent among them as well as those determined with $N2$ and $N3$, but they differ each other (See Table 7). For some regions of UGC 5242, the $N2$ abundances are very small compared with the other values. This could be a problem of the [N\,{\sc ii}] intensity determination or to the particular $N2$ calibration instead of a real dispersion because the $N3$ abundances are very similar to the R$_{23}$ and $P$ values. Since each method has a different caveat, such similarities in the oxygen abundance indicate the robustness of the value. Therefore, even if each of the metallicity values are not very accurate, the average one is probably a very good approximation of the real metallicity of the region. Of course, with all the caution that the use of semi empirical methods involved.  In this sense, the reader has to take into account that the uncertainties in the abundances are only due to the uncertainties in the line intensities and that the canonical uncertainties of each method due to the empirical calibrations have not been considered, mainly because they are of the same amount for all the abundance values. 

\subsection{Is there anything hidden inside some H\,{\sc ii} regions?}

The second issue to discussed, already addressed in the previous section, is the fact that there are some H\,{\sc ii} region with odd abundances, in the sense that they are located at the outskirts of the galaxies but their oxygen abundances are quite high. In most of these regions there are interesting features in their spectra,
such as the detection of the [O\,{\sc i}] line or very high values of the [N\,{\sc ii}]/H$\alpha$ or [S\,{\sc ii}]/H$\alpha$ ratios, which might indicate the existence of shocks or a different nature than a pure H\,{\sc ii} region. We are going to explore these regions with the aid of the diagnostic diagrams. The diagnostic diagrams presented by Baldwin et al. (1981) are based on the [O\,{\sc ii}]/[O\,{\sc iii}] ratio. As we have used the intensity of the [O\,{\sc iii}] line to determine the intensity of the [O\,{\sc ii}] line, this ratio must be somehow contaminated and we preferred not to used any of these diagrams. Instead, the diagrams presented by Veilleux \& Osterbrock (1987) are very interesting. 

\begin{figure}
\centering
\includegraphics[width=8cm]{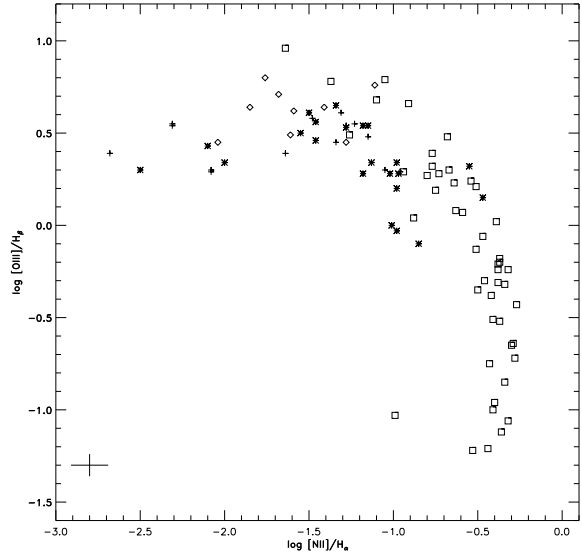}
\caption{The [OIII]/H$\beta$ vs. [NII]/H$\alpha$ diagram for late-type galaxies. The sample of galaxies studied in this investigation are presented as the stars symbols. Squares are the Sm galaxies from McCall et al. (1985), while diamonds are Im galaxies and crosses are dS from the literature. At the bottom-left of the figure the cross represents the typical averaged errorbar of the data-points. } 
\label{fig2}
 \end{figure}

\begin{figure}
\centering
\includegraphics[width=8cm]{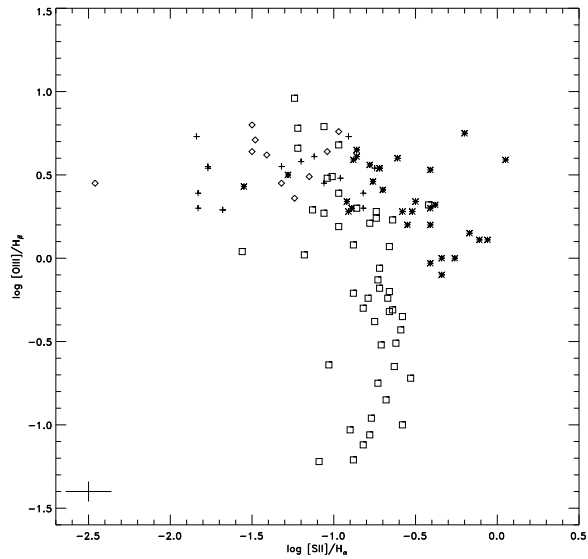}
\caption{The [OIII]/H$\beta$ vs. [SII]/H$\alpha$ diagram for late-type galaxies. Symbols are in figure ~\ref{fig2}. The classical Sm galaxies trace very well the typical H\,{\sc ii} curve while the dS studied in this investigation are located at the right of this curve (See text for details). At the bottom-left part of the figure the cross represents the typical errorbar of the data-points, averaged to the different type of galaxies.} 
\label{fig3}
 \end{figure}

\begin{figure}
\centering
\includegraphics[width=8cm]{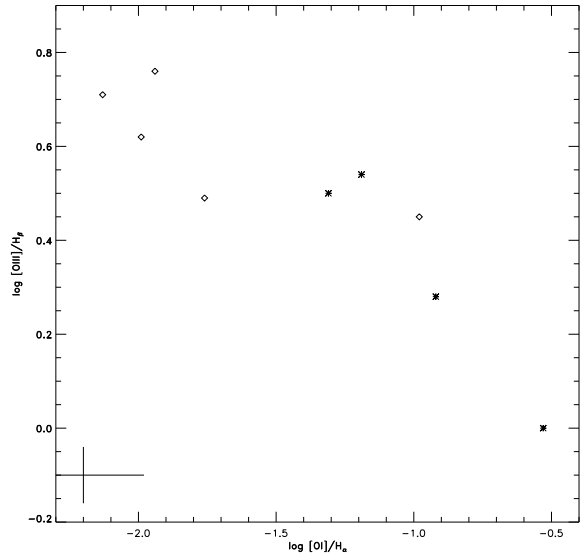}
\caption{The log[OIII]/H$\beta$ vs. [OI]/H$\alpha$ diagram for late-type galaxies. Symbols are in figure ~\ref{fig2}. No information is provided for the classical Sm galaxies and, therefore the H\,{\sc ii} region curve is not defined in this diagram. Dwarf spiral galaxies are at the right while Im are at the left. Although there is a separation between these two types of galaxies, it is not significant considering the errorbars and the small number of data-points. At the bottom-left of the figure the cross represents the typical errorbar of the data-points, considering the different type of galaxies.} 
\label{fig4}
 \end{figure}

In figures ~\ref{fig2}, ~\ref{fig3} and ~\ref{fig4} we plot the log([OIII])/H$\beta$ vs. log([NII])/H$\alpha$, log([SII])/H$\alpha$, and log([OI])/H$\alpha$,  respectively. Along with the data from the galaxies studied in this investigation, we included data from a sample of Im galaxies (Hidalgo-G\'amez et al. 2001; Hidalgo-G\'amez et al. 2001b; Hidalgo-G\'amez \& Olofsson 2002; Hidalgo-G\'amez 2006; Hidalgo-G\'amez 2007; Hidalgo-G\'amez \& Georgiev, in preparation), from the sample of dS published in HG04 and the Sm galaxies from McCall et al. (1985), named Sp in Hidalgo-G\'amez \& Olofsson (2002). Although this is an old investigation, about $35$ years old, their main results have not been superseded by new results and they still hold. 

The H\,{\sc ii} region locus is very well defined by the Sm galaxies from McCall et al. sample for both figures ~\ref{fig2} and ~\ref{fig3}. All but two of the data-points of the dS galaxies studied here  are located to the right of the locus defined by the normal Sm sample in the log([OIII])/H$\beta$ vs. log([NII])/H$\alpha$ diagram, as much of the dS from the literature and the Im sample. Actually, the three samples cannot be distinguished from each other in this diagram, and there is a gap between them and those Sm from McCall et al. (1985). At the same excitation, Sm galaxies have larger values of the [NII])/H$\alpha$ ratio. Following the models from Evans \& Dopita (1985) showed by Veilleux \& Osterbrock (1989), and using the excitation values, the ionization temperature of the star cluster can be inferred, with values between $56,000$ and $45,000$ K for all the H\,{\sc ii} regions of the dS samples, which are larger than those determined in normal Sm galaxies. This might indicate that Sm and dS are not self-similar, but have important and interesting differences in their evolutionary processes, as concluded by HG04. The log([OIII])/H$\beta$ vs. log([NII])/H$\alpha$ diagram is quite interesting in itself because it has also been used recently as a metallicity discriminator (Asari et al. 2007), increasing the abundances towards the left-bottom part of the diagram. In figure ~\ref{fig2} we see that the majority of the H\,{\sc ii} regions studied in this investigation have oxygen abundances quite similar to the Im galaxies, lower than $8.4$ dex. This result is in agreement with those obtained in the previous section. 

Concerning the log([OIII])/H$\beta$ vs. log([SII])/H$\alpha$ diagram (figure \ref{fig3}), all the data-points from the dS studied here are located towards the right-upper part of diagram, indicating a larger value of the [SII]/H$\alpha$ ratio than the rest of the galaxies for the same excitation value. Most of them are UGC 6205 regions. As previously discussed, the [SII]/H$\alpha$ ratio in this galaxy is very high, maybe due to shock waves induced by the spiral pattern. It is important to say that UGC 6205 is the galaxy  with the best defined spiral pattern of the sample studied here. It is also interesting to notice that the data from the dS galaxies are located in a small part of the diagram, contrary to the normal Sm galaxies which fill all the H\,{\sc ii} regions locus. The former are also, in general, separated from the Im galaxies. It might be also interesting to consider the locations of the H\,{\sc ii} regions of our sample in Dopita \& Sutherland's diagrams (1995). Eleven of the H\,{\sc ii} regions are located inside the shocked region of the log([O\,{\sc iii}]/H$\beta$) vs log([S\,{\sc ii}]/H$\alpha$) diagram (their figure 2a). This is in agreement with figure ~\ref{fig3} here, where at least $13$ regions in our sample are located to the left of the H\,{\sc ii} regions locus. When only the high S/N regions are considered, there are no real differences in this diagram. They lie in the same part of the diagram as the low S/N regions, and only few of the latter are located separately, at log([OIII])/H$\beta$ $\approx$ $0$ and log([SII])/H$\alpha$ between $0.0$ and $0.5$. The log([OIII])/H$\beta$ vs. log([NII])/H$\alpha$ diagram from Dopita \& Sutherland (1995) is not very useful here because it is based on very metallic objects and all the regions in our sample but two lay outside the range. These two regions are U05a1 and U77b1, both having a large oxygen abundance. 

Finally, it can be considered the log([OIII])/H$\beta$ vs. log([OI])/H$\alpha$ diagram. There is no information on the [O\,{\sc i}]$\lambda$6300 line intensity for Sm galaxies (see McCall et al. 1985) and therefore, the H\,{\sc ii} region locus is not very well defined. Actually, there are only nine regions in this diagram, five Im and four dS, which is a very small sample for any conclusion. However, it can be seen that there is a single region (U96a2) which is located very likely inside the shocked region, while other two (U96a3 and U77a1) are at the border of it. The only conclusion that could be obtained from figure ~\ref{fig4} is that there is a quite separation between dS and Im, having the former larger values of the [O\,{\sc i}]/H$\alpha$ ratio.  

\subsection{The role of nitrogen}

As an interesting aside, we will discuss the log(N/O) vs. 12+log(O/H) diagram. This is considered as an evolutionary diagram in the sense that oxygen is released into the ISM at an earlier phase than nitrogen, and therefore this ratio is smaller for recent events of star-formation while as the region aged, the log(N/O) becomes larger (Pilyugin 1992). Even without the [O\,{\sc iii}]$\lambda$4363 line, the log(N/O) ratio can be obtained (Lee et al. 2003). An electronic temperature of the high-ionization zone (T$_e^{++}$) is determined from the relation between this parameter and the oxygen abundances (e.g. figure 7 at McGaugh 1991). With this value, equations in Table 1 at Vila-Costas \& Edmunds (1993) can be used in order to obtain an electronic temperature of the low-ionization zone (T$_e^{+}$). Finally, the ratio is determined using equation 9 in Pagel et. al (1992). It has to be said that no correction has been applied to the log (N$^+$/O$^+$) ratio.

It is interesting to notice that the log(N/O) determined with the standard method for U77a2, as stated in Table 8, is much lower than the value determined as previously explained, $-2.1$ dex compared to $-1.8$ dex. Several explanations could be for it: the large value of the T$_e$ obtained, the low intensity of the [OII]$\lambda$3727\AA~ line or the low intensity of the [NII]$\lambda$6583\AA~ line. Actually, the nitrogen abundance from Table 8 is also very low, which might indicates that the last possibility is the most likely. 

\begin{figure}
\centering
\includegraphics[width=8cm]{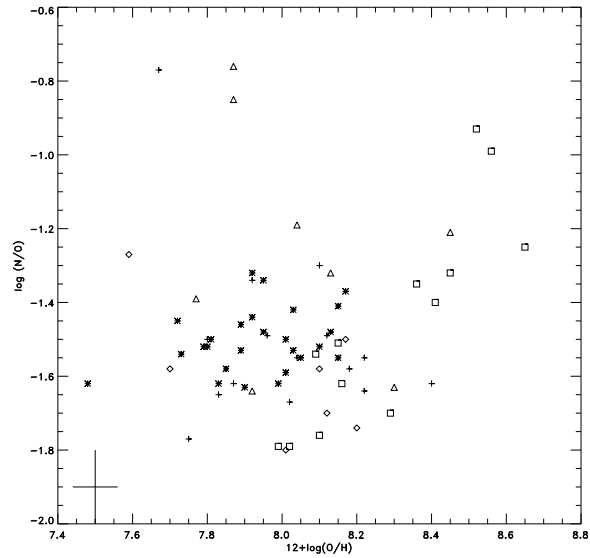}
\caption{The log(N/O) vs. 12+log(O/H) diagram for late-type galaxies. Stars are BCG, crosses LSBGs, diamonds dwarf irregulars, triangle are late-type spirals and squares represent the galaxies studied in this investigation. At the bottom-left part of the figure the cross represents the typical errorbar of the data-points.} 
\label{fig5}
 \end{figure}

From the data presented in this investigation, a value of this ratio is determined for $13$ regions. We did not consider the values obtained with the VISTA software because all the log(N/O) are very odd, with values very low or very high. This diagram is shown in figure ~\ref{fig5} for all the regions studied here (squares) along with a sample of Blue Compact Galaxies (stars), Low-surface Brightness (pluses), dwarf irregular (diamonds) and normal spirals (triangles) from Hidalgo-G\'amez \& Olofsson (2002) and references therein. It is claimed that nitrogen is a secondary element in spiral galaxies (e.g. D\'{\i}az et al. 1991), being related with the oxygen abundance, while it is primary in irregular, showing a flat relation with the oxygen content. This is not very much the situation in this diagram (See figure ~\ref{fig5}). As expected, both LSBG and BCG show a flat relation, considering the large uncertainties associated with this ratio, without galaxies with metallicities larger than $8.4$ dex and $8.2$ dex and $-1.6$ dex the lowest value of the log(N/O). Contrary to their expected behaviour, normal-size late-type spirals (Sm) have a very large dispersion, mostly a scatter diagram, with data-points in the low-metallicity and high log(N/O)region (at the left of the figure) and only two of the galaxies have metallicities larger than $8.2$ dex. Dwarf spiral galaxies follow the trend in this diagram if nitrogen is a secondary element for abundances larger than $8.3$ dex, while for lower abundances the log(N/O) seems to follow a flat relation at a lower value than those for BGC and LSBG. It is also interesting the rising of the trend at very low metallicity. Only one out of $10$ regions with metallicities lower than $7.8$ dex has a value of log(N/O) lower than $-1.6$ while $10$ out of $43$ regions have this value for metallicities between $7.8$ and $8.2$ dex. Most of those regions are not in BCG, therefore this result is not in contradiction with those obtained by Izotov et al. ({\bf ????}), where BCG at low metallicity has a very constant value of the log(N/O). In any case, the dispersion for any type of galaxies at low metallicity is very large, at least $0.3$ dex and a large sample of late-type galaxies is needed to confirm such rising. 

Then, it can be concluded that the secondary origin for nitrogen is related to the oxygen abundances only and not to the morphological type because there are H\,{\sc ii} regions in spiral galaxies with log(N/O) lower than $-1.5$. Then, it cannot be said that nitrogen in spiral galaxies is always secondary, but only in those H\,{\sc ii} regions of high abundance. 

\subsection{Comparison with other investigations}

No one of the H\,{\sc ii} regions studied here have a real high value of the oxygen abundance, as in other spiral galaxies (e.g. up to $9.4$ dex for galaxies in the Virgo Cluster in Skillman et al. 1996). Therefore, it is important to see if the abundances values obtained here are typical of small, late-type spiral galaxies or, on the contrary, they are very different to the previous values determined. 
 
As said in the introduction, there are already 17 dS galaxies with abundance determinations, although most of them were not classified as dS by the authors. In table 4 of HG04 the oxygen abundance of nine of these galaxies were presented. The other eight galaxies have been studied recently. They are presented in Table 9. 
For most of the galaxies, the abundances were not determined with the standard method, but with any of the semi empirical methods, as in the present study. Only for two galaxies (NGC 2552 and UGCA 294) more than one of those methods were used in the determination of the oxygen abundances. And other three galaxies (NGC 5666, UGC 7861 and NGC 625) have abundances values for several H\,{\sc ii} regions.

In HG04 it is concluded, based on nine dS galaxies, that them and Im have very similar oxygen abundance values although normal spiral have higher abundances (Dutil 1998). This result is confirmed with the new abundances published, where the largest value is slightly sub solar, $8.59$ dex in NGC 3510 (Kewley et al. 2005) and the lowest one is $7.69$ dex in NGC 625 (Saviane et al. 2008). The metallicity range encompassed by them is very similar to the one of the present study as well as to the metallicity values of Im. Only three of the H\,{\sc ii} regions studied here have a large value of the oxygen abundance, all of them in UGC 6205, and only U96a1 have a lower metallicity. Two H\,{\sc ii} regions in DDO 204 have very low metallicities, of $7.5$ dex and $7.1$ dex, mainly because of the very low value of the [OII]$\lambda$3727 \AA~ intensities (HG04). When equation 1 is used to calculated a new intensity for this line, values of $8.08$ dex and $7.91$ are obtained, which are similar to the rest of the abundances in dS galaxies.  In conclusion, it can be said that dS galaxies have lower abundances than other type of spiral galaxies, in agreement with the observed trend (Zaritsky et al. 1994). Also, the abundances presented here are in fully agreement with those previously reported.

\section{Conclusions}

The oxygen abundances have been determined for a small sample of dwarf spiral galaxies. The auroral lines were detected in only one out of 29 H\,{\sc ii} regions studied. Therefore, the semi empirical methods were used for the abundances determinations in the rest of the regions. For those H\,{\sc ii} regions where the nitrogen line was detected a total of four methods were used. For those regions where the nitrogen line were absent the abundances were determined using only two methods. For all the regions studied, a weighted average abundance is determined based on all the determinations. 

The majority of the regions have a sub solar abundances, with values between $8.5$ and $7.8$ dex, in agreement with the values obtained to the other types of late galaxies. Only the central part of UGC 6205 have more metallic regions and it is the most metallic galaxy in the sample. The other three galaxies are of low metallicity, with under solar abundances in all their H\,{\sc ii} regions, even the most internal ones.  Actually, the most external region of UGC 5296 seems to have a very low abundance, of only $7.4$ dex, which might be one of the lowest abundance value measured in a spiral galaxy so far. Three of the regions studied here have, probably, a PN embedded because the high [N\,{\sc ii}]/H$\alpha$ ratio while another one might have shocks induced by the spiral arms. This is suggested by the diagnostic diagram and the shock models by Dopita \& Sutherland (1995). Moreover, the galaxies in the sample studied here are located in all the diagnostic diagram closer to the irregular galaxies than to the canonical Sm galaxies, indicating that there might be two different types of Sm galaxies: the dwarf Sm galaxies and the normal-size ones as indicated by other investigations (e.g. Reyes-P\'erez \& Hidalgo-G\'amez 2008). 
   
Concerning the log(N/O) ratio, not all the H\,{\sc ii} regions in spiral galaxies show a secondary behaviour, but only those with metallicity larger than $8.2$ dex.The rest of the regions show a flat relation with the oxygen abundance. A larger sample of dS galaxies is needed to confirm these results (Moranchel-Basurto \& Hidalgo-G\'amez, in preparation). 

\begin{acknowledgements}
The authors thank an anonymous referee who asked the right questions to improved the manuscript. H. Casta\~neda and F.J. S\'anchez-Salcedo are also thanked for a carefully reading of the manuscript. A.M.H-G. thanks J.M. V\,{\'i}lchez for very 
interesting discussions. A.M.H-G.~ is indebted to Felipe Montalvo and Michael Richer at OAN-SPM for their help during the acquisition of the data. This investigation is part of the Master thesis of Daniel Ram\'{\i}rez Fuentes. This investigation was supported by CONACyT CB2006-60526 and SIP20100225. This research has made use of the NASA/IPAC Extragalactic Database (NED) which is operated by the Jet Propulsion Laboratory, California Institute of Technology, under contract with the National Aeronautics and Space Administration. 

\end{acknowledgements}

\begin{table}
\caption[]{Characteristics of the galaxies in the sample presented here. The name of the galaxies is given in column 1 while column 2 shows the morphological type from HG04. The optical radius and the absolute magnitude in the B band, both from HG04, are presented in columns 3 and 4, respectively. The gas mass in 10$^9$ M$_{\odot}$ is shown in column 5 and the surface density (in M$_{\odot}$ per square parsec) in column 6.}
\vspace{0.05cm}
\begin{center}
\begin{tabular}{c c c c c c} 
\hline
{Galaxy} & {Type} & { r(kpc)} & {M$_B$} & {M(HI)} & {$\rho_s$} \\
\hline
UGC 5296 & S9    & 3.02 & -15.08 & 0.29 & 10.12\\
UGC 6205 & Sm    & 4.99 & -16.53 & 0.31 & 3.96\\
UGC 5242 & SBS9  & 4.86 & -16.60 & 0.80 & 10.78\\
UGC 6377 & dS    &  -   & -15.45 & - &  -\\
\hline
\end{tabular}
\end{center}
\end{table}

\begin{table}
\caption[]{Log of Observations. The galaxies observed are presented in column 1. Column 2 shows the date of observation while slit positions are presented in column 3. The coordinates (right ascension in column 4 and declination in column 5) for each slit position are giving. The total integration time (in seconds) is in column 6 and the maximum air mass in column 7. Column 8 shows the angle of the rotator, where 0° means to  }
\vspace{0.05cm}
\begin{center}
\begin{tabular}{c c c c c c c}
\hline
{Galaxy}  & {Date} & {Slit position} & {$\alpha$} & {$\delta$} & {Integr. Time (s)} & {Air mass} \\
\hline 
UGC 5296  & 10/03/02  & a  & 09 53 19  & 58 27 52  &  3600  &  1.2   \\
UGC 6205  & 10/03/02  & a  & 11 10 04  & 46 04 50  &  2700  &  1.05    \\
          & 10/03/02  & b  & 11 09 50  & 46 04 55  &  3600  & 1.1     \\
          & 10/03/02  & c  & 11 10 02  & 46 05 15  &  3300  &  1.4 \\
UGC 5242  & 11/03/02  & a  & 09 47 00  & 00 57 00  &   900  & 1.2 \\
          & 11/03/02  & b  & 09 47 00  & 00 57 00  &   900  &  1.2 \\
          & 11/03/02  & c  & 09 47 00  & 00 57 00  &  3600  &  1.2 \\
          & 11/03/02  & d  & 09 47 00  & 00 57 00  &  3600  & 1.2  \\
          & 11/03/02  & e  & 09 47 00  & 00 57 00  &  3600  &  1.3  \\
UGC 6377  & 11/03/02  & a  & 11 27 52  & 41 13 48  &  4200  &  1.3  \\
          & 11/03/02  & b  & 11 27 52  & 41 13 48  &  1800  &  1.5 \\
          & 11/03/02  & c  & 11 27 52  & 41 13 48  &  1500  &  1.7  \\
\hline
\end{tabular}
\end{center}
\end{table} 

\clearpage

\begin{deluxetable}{cccccccccccc}
\rotate
\tablecolumns{12}
\tabletypesize{\footnotesize}
\tablewidth{0pc} 
\tablecaption{Extinction-corrected line intensities, obtained using the sotware ALICE-MIDAS for UGC 6205. The intensities of the lines along with their uncertainties are shown in columns from 2 to 12 for each of the regions detected in this galaxy. The extinction coefficient in the recombination line H$\alpha$ and the S/N ratio in the [O\,{\sc iii}]$\lambda$5007~\AA~ for each region are also given at the end of each column.}
\tablehead{  
\colhead{Region} & 
\colhead{U05a1} & 
\colhead{U05a2} & 
\colhead{U05a3} & 
\colhead{U05b1} & 
\colhead{U05b2} & 
\colhead{U05b3} & 
\colhead{U05b4} & 
\colhead{U05b5} & 
\colhead{U05c1} &
\colhead{U05c2} &
\colhead{U05c3} \\
}
\startdata
$\lbrack$OII$\rbrack$$\lambda$3727 & 1.7$\pm$0.6 & 1.5$\pm$0.5 & 1.5$\pm$0.5 & 2.0$\pm$0.7 & 2.6$\pm$0.9 & 1.4$\pm$0.5 & 2.6$\pm$0.9 & 1.7$\pm$0.6  & 0.8$\pm$0.3 & 2.4$\pm$0.8 & 2.7$\pm$0.9   \\       
H$\beta$ &  1.0$\pm$0.2 & 1.0$\pm$0.2 & 1.0$\pm$0.2 & 1.0$\pm$0.2 & 1.0$\pm$0.3 & 1.0$\pm$0.3 & 1.0$\pm$0.3 & 1.0$\pm$0.2 & 1.0$\pm$0.3 & 1.0$\pm$0.3 & 1.0$\pm$0.2 \\
$\lbrack$OIII$\rbrack$$\lambda$4959 & 0.7$\pm$0.2 & 0.5$\pm$0.1 & 0.4$\pm$0.1 & 0.5$\pm$0.1 & 1.1$\pm$0.3 & 1.8$\pm$0.6  & 1.1$\pm$0.4  & 0.5$\pm$0.1  & 0.3$\pm$0.1 & 1.3$\pm$0.4 & 0.6$\pm$0.1  \\
$\lbrack$OIII$\rbrack$$\lambda$5007 & 1.4$\pm$0.4 & 1.3$\pm$0.3 & 1.3$\pm$0.3 & 1.6$\pm$0.4 & 3.4$\pm$1 & 5.6$\pm$2 & 3.5$\pm$1 & 1.4$\pm$0.4 & 1$\pm$0.3 & 3.9$\pm$1 & 2.2$\pm$0.4 \\
He~I$\lambda$5875 &- & - & - & - & - & - & 0.11$\pm$0.04 & - & - & - & -\\
$\lbrack$OI$\rbrack$$\lambda$6300 & - & - & - & - & - & - & 0.14$\pm$0.6 & - & - & - & -\\
H$\alpha$ & 2.37$\pm$0.5 & 2.7$\pm$0.7 & 2.4$\pm$0.6 & 2.04$\pm$0.5 & 2.86$\pm$0.9 & 2.86$\pm$1 & 2.14$\pm$0.5 & 2.5$\pm$0.5 & 2.2$\pm$0.6 & 2.86$\pm$0.9 & 2.86$\pm$0.5 \\
$\lbrack$NII$\rbrack$$\lambda$6583 & 0.8$\pm$0.4 & - & - & - & 0.15$\pm$0.09 & - & 0.14$\pm$0.07 & -  & - & - & 0.3$\pm$0.08 \\
$\lbrack$SII$\rbrack$$\lambda$6716+6736 & - & 2.1$\pm$0.7 & 2.1$\pm$0.6 & - & 1.1$\pm$0.4 & 1.8$\pm$0.8 & 0.5$\pm$0.2 & 1.7$\pm$0.7 & 1.2$\pm$0.3 & 3.2$\pm$1 & 0.9$\pm$0.2 \\
  & & & & & & & & & & & \\
C$\beta$(H$\alpha$) & - & - & - & - & 1.4$\pm$0.4 & 2.1$\pm$0.1 & - & - & - & 1.3$\pm$0.06 & 1.2$\pm$0.01 \\
S/N & 11 & 9 & 13 & 7 & 15 & 14 & 20 & 11 & 7 & 5 & 10\\
\enddata
\end{deluxetable}

\begin{table}
\scriptsize
\caption[]{Extinction-corrected line intensities, obtained using the software ALICE-MIDAS for UGC 6377. The intensities of the lines along with their uncertainties are shown in columns 2 to 10 for each of the regions detected in this galaxy. The intensities of the spectral lines from the SLOAN data-base are given in column 11. The extinction coefficient in the recombination line H$\alpha$ and the S/N ratio in the [O\,{\sc iii}]$\lambda$5007~\AA~ for each region are also given at the end of each column. }
\vspace{0.05cm}
\begin{center}
\begin{tabular}{c c c c c c c c c c c}
\hline
{line} & {u77a1} & {u77a2} & {u77a3} & {u77b1} & {u77b2} & {u77b3} & {u77c1} & {u77c2} & {u77c3} & {SLOAN}\\
\hline
$\lbrack$OII$\rbrack$$\lambda$3727 & 2.0$\pm$0.6 & 2.8$\pm$1 & 2.4$\pm$0.8 & 2.6$\pm$0.9 & 2.4$\pm$0.8 & 2.8$\pm$0.4 & 2.7$\pm$0.9 & 2.6$\pm$0.9 & 2.5$\pm$0.4 & 2.7$\pm$0.9\\ 
H$\gamma$ & - & 0.17$\pm$0.03 & - & - & - & - & - & 0.3$\pm$0.1 & - & -\\
$\lbrack$OIII$\rbrack$$\lambda$4363  & - &  0.10$\pm$0.04 & - & - & - & - & - & - & - & -\\ 
HeI$\lambda$4778 & - & 0.10$\pm$0.03 & - & - & - & - & - & - & - & -\\ 
H$\beta$ & 1.0$\pm$0.2 & 1.0$\pm$0.04 & 1.0$\pm$0.2 & 1.0$\pm$0.4 & 1.0$\pm$0.7 & 1.0$\pm$0.5 & 1.0$\pm$0.4 & 1.0$\pm$0.7 & 1.0$\pm$.1 & 1$\pm$0.03\\
HeI$\lambda$49 & - & 0.14$\pm$0.06 & - & - & - & - & - & - & - & -\\  
$\lbrack$OIII$\rbrack$$\lambda$4959  & 0.5$\pm$0.1 & 1.1$\pm$0.07 & 1.1$\pm$0.5 & 0.7$\pm$0.4 & 1.3$\pm$1 & 1.3$\pm$1 & 1.1$\pm$0.5 & 1.2$\pm$1 & 0.6$\pm$0.4 & 0.4$\pm$0.2\\
$\lbrack$OIII$\rbrack$$\lambda$5007 & 1.6$\pm$0.2 & 2.9$\pm$0.2 & 3.9$\pm$1 & 2.1$\pm$0.8 & 4$\pm$3 & 2.5$\pm$1 & 3.1$\pm$1 & 3.5$\pm$3 & 1.9$\pm$1 & 2.1$\pm$0.2\\  
$\lbrack$NI$\rbrack$$\lambda$5200  & - & 0.13$\pm$0.05 & - & - & - & - & - & - & - & -\\  
HeI$\lambda$5875 & - &  0.17$\pm$0.05 & - & - & - & - & - & - & - & -\\ 
$\lbrack$OI$\rbrack$$\lambda$6300  & 0.3$\pm$0.1 & - & - & - & - & - & - & - & -\\ 
H$\alpha$ & 2.86$\pm$0.4 & 2.82$\pm$0.1 & 2.86$\pm$0.6 & 2.86$\pm$1 & 2.86$\pm$2 & 2.07$\pm$1 & 2.86$\pm$1 & 2.86$\pm$2 & 2.25$\pm$1 & 2.86$\pm$0.2\\ 
$\lbrack$NII$\rbrack$$\lambda$6583 & 0.3$\pm$0.1 & 0.10$\pm$0.03 & - & 0.8$\pm$0.5 & - & - & - & 0.20$\pm$0.03 & - & -\\
$\lbrack$SII$\rbrack$$\lambda$6716+6736 & 0.8$\pm$0.2 & 0.50$\pm$0.05 & 0.38$\pm$0.1 & 1.2$\pm$0.5 & 0.7$\pm$0.5 & - & - & 0.55$\pm$0.06 & - & -\\
                     &     & & & & & & & & \\
C$\beta$(H$\alpha$)  & 0.54$\pm$0.1 & - & 0.41$\pm$0.1 & 0.3$\pm$0.1 & 2.9$\pm$2 & - & 0.8$\pm$0.4 & 0.3$\pm$0.03 & - & \\
S/N & 20 & 45 & 5 & 15 & 7 & 4 & 4 & 13 & 3 \\
\hline 
\end{tabular}
\end{center}
\end{table}

\begin{table}
\caption[]{Extinction-corrected line intensities, obtained using the sotware ALICE-MIDAS (first row) and VISTA (second row) for UGC 5296. The intensities of the lines along with their uncertainties are shown in columns 2, 3, 4, and 5 for each of the regions detected in this galaxy, while column 1 shows the spectral lines detected. The extinction coefficient in the recombination line H$\alpha$ and the S/N ratio in the [O\,{\sc iii}]$\lambda$5007~\AA~ for each region are also given at the end of each column. }
\vspace{0.05cm}
\begin{center}
\begin{tabular}{c c c c c}
\hline
{line}  & {u96a1} & {u96a2} & {u96a3} & {u96a4}   \\
\hline 
$\lbrack$OII$\rbrack$$\lambda$3727      & 1.22$\pm$0.4  & 0.92$\pm$0.3  & 2.41$\pm$0.8  & 2.58$\pm$0.9    \\
                                        & 0.50$\pm$0.2  & 0.76$\pm$0.3  & 2.33$\pm$0.8  & 2.50$\pm$0.9     \\
H$\gamma$                               & -             &   -           & -             &  0.47$\pm$0.1       \\ 
                                        & -             &   -           &  -            &   -                \\
H$\beta$                                & 1.0$\pm$0.4   & 1.0$\pm$0.1   & 1.0$\pm$0.1   & 1.0$\pm$0.2  \\
$\lbrack$OIII$\rbrack$$\lambda$4959     & 0.41$\pm$0.2  & 0.37$\pm$0.04 & 0.73$\pm$0.08 & 0.4$\pm$0.2    \\
                                        & 0.28$\pm$0.2  & 0.38$\pm$0.04 & 0.64$\pm$0.07 & 0.65$\pm$0.2    \\
$\lbrack$OIII$\rbrack$$\lambda$5007     & 1.2$\pm$0.5   & 1.0$\pm$0.1   & 1.9$\pm$0.1   & 2.0$\pm$0.4   \\
                                        & 0.80$\pm$0.05 & 0.94$\pm$0.1  & 1.8$\pm$0.1   & 1.9$\pm$0.4   \\
$\lbrack$OI$\rbrack$$\lambda$6300       &  -            & 0.85$\pm$0.3  & 0.33$\pm$0.1  &   -          \\
                                        &  -            &     -         &    -          &  -  \\
H$\alpha$                               & 2.1$\pm$0.8   & 2.86$\pm$0.3  & 2.74$\pm$0.2  & 2.86$\pm$0.7    \\
                                        & 2.86$\pm$1    & 2.86$\pm$0.3  & 2.86$\pm$0.2  & 2.86$\pm$0.6  \\
$\lbrack$NII$\rbrack$$\lambda$6583      &   -           & 0.28$\pm$0.05 & 0.29$\pm$0.02 &   -      \\ 
                                        & 0.40$\pm$0.3  & 0.30$\pm$0.05 & 0.27$\pm$0.02 & 0.19$\pm$0.08  \\
$\lbrack$SII$\rbrack$$\lambda$6716+6736 &      -        & 1.3$\pm$0.1   & 0.72$\pm$0.05 & 0.37$\pm$0.1  \\
                                        &  1.3$\pm$0.8  & 1.1$\pm$0.1   & 0.87$\pm$0.1  & 0.35$\pm$0.1 \\
                                        &               &               &               &              \\  
C$\beta$(H$\alpha$)                     &    -          & 0.46$\pm$0.03 &  -            & 0.49$\pm$0.09 \\
                                        &  0.77$\pm$0.3 & 1.65$\pm$0.1  & 0.80$\pm$0.07 & 2.2$\pm$0.4     \\
S/N                                     & 3             &   7           &   12          &  6            \\
\hline
\end{tabular}
\end{center}
\end{table}

\begin{table}
\caption[]{Extinction-corrected line intensities, obtained using the sotware ALICE-MIDAS (first row) and VISTA (second row) for UGC 5242. The intensities of the lines along with their uncertainties are shown in columns 2, 3, 4, 5 and 6 for each of the regions detected in this galaxy. The intensities of the spectral lines from the SLOAN data-base are given in column 7. The extinction coefficient in the recombination line H$\alpha$ and the S/N ratio in the [O\,{\sc iii}]$\lambda$5007~\AA~ for each region are also given at the end of each column. }
\vspace{0.05cm}
\begin{center}
\begin{tabular}{c c c c c c c}
\hline
{line}  & {u42c1} & {u42c2} & {u42d1} & {u42e1} & {u42e2} & {SLOAN} \\
\hline 
$\lbrack$OII$\rbrack$$\lambda$3727      & 2.3$\pm$0.8   & 2.1$\pm$0.9   & 2.5$\pm$0.9   & 2.0$\pm$0.7 & 2.8$\pm$1       & 2.73$\pm$0.9     \\
                                        & 2.8$\pm$1     & 2.5$\pm$0.9   &    -          & 2.3$\pm$0.8   & 2.7$\pm$0.9   &  - \\
H$\gamma$                               & 0.40$\pm$0.05 &   -           & 0.2$\pm$0.02  &    -          &  -            &  -     \\
                                        &        -      &   -           &   -           &     -         &  -            &  -  \\
H$\beta$                                & 1.0$\pm$0.05  & 1.0$\pm$0.5   & 1.0$\pm$0.06  & 1.0$\pm$0.07  & 1.0$\pm$0.1   & 1.0$\pm$0.1 \\
$\lbrack$OIII$\rbrack$$\lambda$4959     & 1.1$\pm$0.08  & 0.53$\pm$0.3  & 1.1$\pm$0.07  & 1.5$\pm$0.1   & 0.9$\pm$0.1   & 0.67$\pm$0.06 \\
                                        & 1.0$\pm$0.07  & 0.68$\pm$0.3  &   -           & 1.5$\pm$0.1   & 0.8$\pm$0.1   &  - \\
$\lbrack$OIII$\rbrack$$\lambda$5007     & 3.2$\pm$0.2   & 1.6$\pm$0.9   & 3.6$\pm$0.2   & 4.5$\pm$0.3   & 2.6$\pm$0.3   & 2.2$\pm$0.1\\
                                        & 2.7$\pm$0.2   & 2.0$\pm$1     &   -           & 4.1$\pm$0.3   & 2.2$\pm$0.3   &  -\\
HeI$\lambda$5875                        & 0.05$\pm$0.02 &   -           & 0.15$\pm$0.04 &  -            &   -           &  - \\
                                        &   -           &   -           &    -          &   -           &    -          &  -  \\
$\lbrack$OI$\rbrack$$\lambda$6300       & 0.14$\pm$0.05 &  -            &   -           &  -            &  -            &  -  \\
                                        &   -           &            -  &       -       &   -           &  -            &  -  \\
$\lbrack$NII$\rbrack$$\lambda$6548      &  -            &   -           & 0.06$\pm$0.03 &  -            &   -           &  - \\
                                        &   -           &   -           &   -           &   -           &   -           &  -\\
H$\alpha$                               & 2.86$\pm$0.1  & 2.57$\pm$1    & 2.86$\pm$0.2  & 2.86$\pm$0.2  & 2.86$\pm$0.3  & 2.86$\pm$0.2  \\
                                        & 2.86$\pm$0.1  & 2.86$\pm$1    &  -            & 2.86$\pm$0.2  & 2.86$\pm$0.3  & - \\
$\lbrack$NII$\rbrack$$\lambda$6583      & 0.08$\pm$0.03 &    -          & 0.10$\pm$0.01 & 0.13$\pm$0.01 &  -            & 0.21$\pm$0.02 \\ 
                                        & 0.02$\pm$0.01 & 0.01$\pm$0.01 &  -            & 0.10$\pm$0.01 & 0.03$\pm$0.01 & -  \\
HeI$\lambda$6877                        &    -          &    -          & 0.04$\pm$0.01 &  -            &   -           &   -   \\
                                        &          -    &      -        &    -          &   -           &   -           &   -  \\
$\lbrack$SII$\rbrack$$\lambda$6716+6736 & 0.15$\pm$0.06 & 1.0$\pm$0.6   & 0.47$\pm$0.03 & 0.39$\pm$0.04 & 0.56$\pm$0.1 &  -\\
                        &  0.08$\pm$0.03 & 1.1$\pm$0.6 &   -            & 0.39$\pm$0.04 & 0.34$\pm$0.04 &  -\\
                        &                &             &                &               &               &   \\
C$\beta$(H$\alpha$)     &  2.51$\pm$0.01     &    -        & 1.14$\pm$0.01     & 1.55$\pm$0.01     & 0.65$\pm$0.02     &  1.2$\pm$0.08  \\
S/N                     & 57             &   4         &   43          &   33          &  20           &   \\ 
\hline
\end{tabular}
\end{center}
\end{table}

\begin{table}\scriptsize
\caption[]{Oxygen abundaces for each of the H\,{\sc ii} regions studied here. The name of the region is given in column 1, while the log(N/O) ratio is given in column 2. The abundances of the semi empirical methods used in this investigation are presented in columns 3 (the R$_{23}$), 4 (the $P$), 4 ($N2$) and 5 ($N3$), along with their uncertainties. Column 7 shows the weighted-average abundance for each region along with the dispersion among the data. }
\vspace{0.05cm}
\begin{center}
\begin{tabular}{c c c c c c c}
\hline
{region} & log(N/O) & {Z$_{23}$} & {Z$_P$} & {Z$_{N2}$} & {Z$_{N3}$} & {Z$_{ave}\pm \sigma$}  \\
\hline
{u05a1} & {-0.3$\pm$0.2} & {8.8$\pm$0.1}  & {8.55$\pm$0.1} & {8.8$\pm$0.1}   & {8.50$\pm$0.01} & {8.6$\pm$0.1}  \\
{u05a2} & { -          } & {8.8$\pm$0.2}  & {8.58$\pm$0.1} & {0.00}          & {0.00}          & {8.7$\pm$0.1}  \\
{u05a3} & {    -}        & {8.8$\pm$0.1}  & {8.60$\pm$0.2} & {0.00}          & {0.00}          & {8.7$\pm$0.1}  \\
{u05b1} & {   -        } & {7.7$\pm$0.2}  & {7.68$\pm$0.2} & {0.00}          & {0.00}          & {7.7$\pm$0.04} \\
{u05b2} & {-1.2$\pm$0.3} & {8.1$\pm$0.1}  & {8.20$\pm$0.1} & {8.2$\pm$0.3}   & {8.1$\pm$0.1}   & {8.2$\pm$0.04} \\
{u05b3} & {     -      } & {8.1$\pm$0.3}  & {7.87$\pm$0.3} & {0.00}          & {0.00}          & {7.9$\pm$0.08} \\
{u05b4} & {-1.3$\pm$0.4} & {8.4$\pm$0.2}  & {8.34$\pm$0.2} & {8.3$\pm$0.2}   & {8.20$\pm$0.03} & {8.3$\pm$0.1}  \\
{u05b5} & {    -       } & {7.6$\pm$0.2}  & {7.5$\pm$0.2}  & {0.00}          & {0.00}          & {7.6$\pm$0.1}  \\
{u05c1} & {     -      } & {8.97$\pm$0.1} & {8.74$\pm$0.1} & {0.00}          & {0.00}          & {8.8$\pm$0.1}  \\ 
{u05c2} & {      -     } & {8.4$\pm$0.2}  & {8.30$\pm$0.2} & {0.00}          & {0.00}          & {8.4$\pm$0.1}  \\
{u05c3} & {-0.9$\pm$0.2} & {8.6$\pm$0.1}  & {8.35$\pm$0.2} & {8.4$\pm$0.2}   & {8.28$\pm$0.06} & {8.4$\pm$0.1}  \\
{}      & {}     & {}     & {}            & {}              & {}              & \\
{u77a1} & {-0.4$\pm$0.6} & {8.7$\pm$0.1}   & {8.48$\pm$0.1} & {8.4$\pm$0.1}   & {8.35$\pm$0.05} & {8.4$\pm$0.1}  \\
{u77a2} & {-1.4$\pm$0.3} & {8.06$\pm$0.07} & {7.85$\pm$0.1} & {8.06$\pm$0.09} & {8.12$\pm$0.03} & {8.1$\pm$0.1}  \\
{u77a3} & {   -        } & {8.1$\pm$0.3}   & {7.97$\pm$0.2} & {0.00}          & {0.00}          & {8.0$\pm$0.03} \\
{u77b1} & {-0.5$\pm$0.7} & {8.6$\pm$0.2}   & {8.37$\pm$0.2} & {8.7$\pm$0.3 }  & {8.46$\pm$0.07} & {8.6$\pm$0.1}  \\
{u77b2} & {     -      } & {8.4$\pm$0.4}   & {8.31$\pm$0.2} & {0.00}          & {0.00}          & {8.3$\pm$0.1}  \\
{u77b3} & {     -      } & {8.0$\pm$0.2}   & {7.95$\pm$0.3} & {0.00}          & {0.00}          & {8.0$\pm$0.01} \\
{u77c1} & {     -      } & {8.1$\pm$0.2}   & {7.97$\pm$0.3} & {0.00}          & {0.00}          & {8.0$\pm$0.02} \\
{u77c2} & {-1.1$\pm$0.3} & {8.1$\pm$0.1}   & {7.99$\pm$0.2} & {8.3$\pm$0.4}   & {8.2$\pm$0.2}   & {8.1$\pm$0.1}  \\
{u77c3} & {    -       } & {7.9$\pm$0.1}   & {7.72$\pm$0.2} & {0.00}          & {0.00}          & {7.8$\pm$0.06} \\
{SLOAN} & {}             & {7.9$\pm$0.1}   & {7.70$\pm$0.1} & {0.00}          & {0.00}          & {7.8$\pm$0.1}  \\
{}      & {}     & {}     &  {}             & {}              & {}              & \\
{u96a1} & {    -       } & {7.5$\pm$0.5}   & {7.37$\pm$0.2} & {0.00}          & {0.00}          & {7.4$\pm$0.04} \\
        & {-0.5$\pm$0.3} & {7.1$\pm$0.1}   & {7.17$\pm$0.2} & {8.5$\pm$0.1}   & {8.49$\pm$0.06} & {8.0$\pm$0.7}  \\
{u96a2} & {-0.5$\pm$0.2} & {8.93$\pm$0.04} & {8.71$\pm$0.1} & {8.4$\pm$0.1}   & {8.40$\pm$0.03} & {8.5$\pm$0.2}\\
        & {-0.4$\pm$0.1} & {8.97$\pm$0.01} & {8.74$\pm$0.1} & {8.4$\pm$0.1}   & {8.40$\pm$0.05} & {8.5$\pm$0.2}  \\
{u96a3} & {-0.9$\pm$0.3} & {8.7$\pm$0.1}   & {8.40$\pm$0.2} & {8.41$\pm$0.04} & {8.33$\pm$0.09} & {8.4$\pm$0.1}  \\
        & {-0.9$\pm$0.3} & {8.7$\pm$0.1}   & {8.43$\pm$0.2} & {8.37$\pm$0.05} & {8.32$\pm$0.01} & {8.4$\pm$0.1} \\
{u96a4} & {    -       } & {7.9$\pm$0.1}   & {7.72$\pm$0.2} & {0.00}          & {0.00}          & {7.8$\pm$0.07} \\
        & {-1.1$\pm$0.3} & {7.8$\pm$0.1}   & {7.73$\pm$0.2} & {8.3$\pm$0.2}   & {8.26$\pm$0.06} & {8.1$\pm$0.2}  \\
{}      &  {}    &  {}    &  {}            &  {}             & {}              & \\ 
{u42c1} & {1-.5$\pm$0.1} & {8.0$\pm$0.1}   & {7.99$\pm$0.2} & {7.9$\pm$0.2}   & {8.05$\pm$0.06} & {8.0$\pm$0.05} \\ 
        & {-2.1$\pm$0.4} & {8.0$\pm$0.1}   & {7.93$\pm$0.2} & {7.5$\pm$0.1}   & {7.90$\pm$0.05} & {7.8$\pm$0.2}  \\
{u42c2} & {    -       } & {7.7$\pm$0.3}   & {7.60$\pm$0.3} & {0.00}          & {0.00}          & {7.6$\pm$0.07} \\
        & {-2.4$\pm$0.6} & {7.8$\pm$0.3}   & {7.74$\pm$0.3} & {7.3$\pm$0.4}   & {7.8$\pm$0.1}   & {7.8$\pm$0.2}  \\
{u42d1} & {-1.4$\pm$0.2} & {8.1$\pm$0.1}   & {7.98$\pm$0.2} & {8.05$\pm$0.06} & {8.2$\pm$0.1}   & {8.1$\pm$0.1}  \\
{u42e1} & {-1.2$\pm$0.2} & {8.1$\pm$0.1}   & {7.98$\pm$0.2} & {8.15$\pm$0.04} & {8.10$\pm$0}    & {8.1$\pm$0.06} \\
        & {-1.4$\pm$0.2} & {8.0$\pm$0.1}   & {7.99$\pm$0.2} & {8.06$\pm$0.05} & {8.07$\pm$0.01} & {8.0$\pm$0.03} \\
{u42e2} & {    -       } & {8.0$\pm$0.1}   & {7.91$\pm$0.3} & {0.00}          & {0.00}          & {7.9$\pm$0.03} \\
        & {-1.9$\pm$0.3} & {7.9$\pm$0.1}   & {7.82$\pm$0.3} & {7.7$\pm$0.1}   & {7.99$\pm$0.03} & {7.8$\pm$0.1}  \\
{SLOAN} & {-1.1$\pm$0.2} & {7.9$\pm$0.1}   & {7.81$\pm$0.2} & {8.29$\pm$0.05} & {8.20$\pm$0.07} & {8.0$\pm$0.2}  \\ 
\hline
\end{tabular}
\end{center}
\end{table}

\begin{table}
\caption[]{Electron temperatures, oxygen and nitrogen abundances for region U77a2 in UGC 6377, determined with the standard method. }
\vspace{0.05cm}
\begin{center}
\begin{tabular}{c c }
\hline
{Parameter} & {Value}\\
\hline
Te(O$^{++}$)  & 20,500$\pm$500 (K)\\
Te(O$^+$)     &  16,700$\pm$50 (K)\\
12+log(O/H)   & 7.8$\pm$0.15 \\
12+log(N/H)   & 5.6$\pm$0.1 \\
log(N/O)      & -2.1$\pm$0.3 \\
\hline
\end{tabular}
\end{center}
\end{table}

\begin{table}
\caption[]{Oxygen abundances of dS galaxies, previously reported in the literature. The name of the galaxy is presented in column 1 while the abundance value is given in column 2. References are given in column 3: S08 stands for Saviane et al. (2008), PT07 for Pilyungin \& Thuan (2007), C09 for Crox et al. (2009), KJG05 for Kewley et al. (2005), GP07 for Gil de Paz et al. (2007) and SKLC05 for Shi et al. (2005). }
\vspace{0.05cm}
\begin{center}
\begin{tabular}{c c c}
\hline
{Galaxy} & {Value} & {Reference}\\
\hline
NGC 625   & 7.69-8.37 & S08\\
NGC 2552  & 8.18-8.44 & PT07 \\
UGC 5666  & 7.83-7.99 & C09 \\
UGC 5692  & 7.95      & C09 \\
NGC 3510  & 8.59      & KJG05 \\
UGC 7861  & 7.79-8.53 & GP07 \\
UGCA 294  & 8.21-8.43 & SKLC05\\
UGCA 442  & 7.72      & S08 \\
\hline
\end{tabular}
\end{center}
\end{table}

\end{document}